# BBU effect in an ERL-FEL two-purpose test facility

CUI Xiao-Hao(崔小昊), JIAO Yi(焦毅), WANG Jiu-Qing(王九庆)[1)], WANG Shu-Hong(王书鸿)

Institute of High Energy Physics, Chinese Academy of Sciences, Beijing 100049, China

**Abstract**: Both the Energy Recovery Linac (ERL) and Free Electron laser (FEL) are considered to be candidates of the fourth generation light source. It is proposed to combine FEL into an ERL facility to integrate the advantages of both ERL and FEL, and to realize a compact two-purpose light source. A test facility to verify this principle is being designed at the Institute of High Energy Physics, Beijing. One main concern is the beam breakup (BBU) instability which limits the available beam current. To this end, we developed a numerical simulation code to calculate the BBU threshold, which is found to have only a small reduction even in a high-FEL-bunch-charge operation mode, compared with that in the case with ERL bunches only. However, even with ERL beam current far below BBU threshold, we observed a fluctuation of the central orbit of the ERL bunches in the presence of FEL beam. We then present a physical model of BBU and understand the mechanism of the orbit-fluctuation in an ERL-FEL two-purpose machine. We found that by choosing an appropriate FEL bunch repetition rate, the central orbit fluctuation amplitude can be well controlled.

**Key words:**   ERL, FEL, two-purpose machine, BBU

**PACS** 29.27.Bd, 29.20.Ej, 41.60.Ap

## 1. Introduction

Energy recovery linac (ERL), as one candidate of the fourth generation light source, has the potential to provide high repetition rate, low emittance electron beam and to deliver x rays with high average brilliance (see, e.g. [1]). On the other hand, in spite of low repetition rate, free electron laser (FEL) has shown remarkable success in delivering x rays with extremely high peak brilliance (see, e.g. [2]), and therefore has received much attention worldwide. Since both ERL and FEL are based on linac technologies, it is possible to combine FEL into an ERL facility, resulting in a compact two-purpose light source [3]. Based on this point, we have proposed to build an ERL test facility (ERL-TF) accommodating FEL at the Institute of High Energy Physics (IHEP), Beijing, and have been studying the related beam physics in the past few years [4-5].

Figure 1 shows the schematic layout of the facility. Two series of electron bunches (5 MeV vs. 20 MeV, 130 MHz vs. 10.83 MHz) are injected from separate beam-lines into the main linac consisting of two 9-cell 1.3 GHz superconducting (SC) cavities, and are accelerated to 35 MeV and 50 MeV, for ERL and FEL purposes, respectively. In the main linac, an FEL bunch passes the SC cavities at a different phase of the electric field compared with the ERL bunch. The time structure of the electron bunches in the SC cavities is shown in Fig. 2. The main parameters of the facility are listed in Table 1.

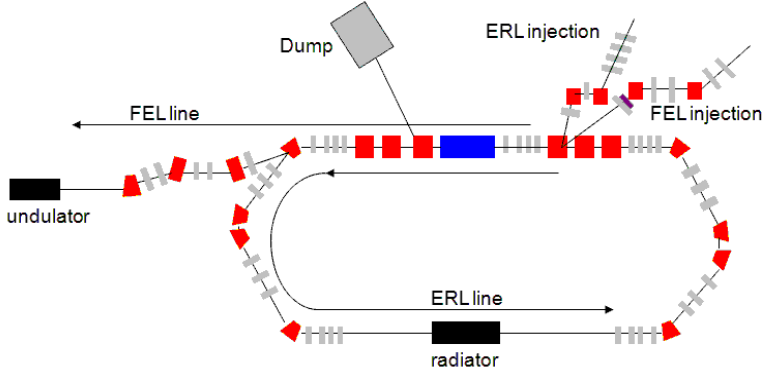

Fig. 1. Layout of the ERL-FEL test-facility

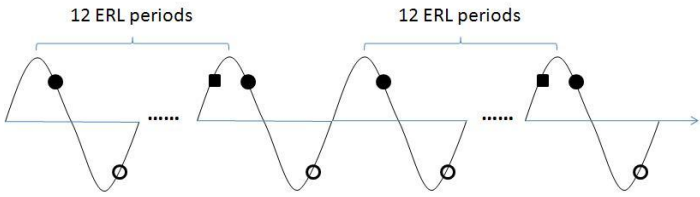

Fig. 2. Time structure of the injected ERL bunches (black dot), recirculated ERL bunches (circle), and FEL bunches (square).

Table 1. The main parameters of the IHEP ERL-TF

| Parameter | ERL | FEL | Unit |
|---|---|---|---|
| Injection energy | 5 | 20 | MeV |
| Max Energy | 35 | 50 | MeV |
| Bunch Charge | 77 | 100 | pC |
| Bunch spacing | 0.77 | 93 | ns |

One important limitation on the available beam current in an ERL is the (transverse) beam breakup (BBU) instability, which has been well studied in the past few years [6-7]. In the simplest model of BBU in ERL, only one cavity with one dipole higher order mode (HOM) is assumed. Electron bunches are injected into the cavity, accelerated, and then recirculated to pass through the cavity during the decelerating phase before they are ejected into the beam dump. Considering one dipole HOM excited in the cavity, a bunch passing through the cavity during the first time experiences a transverse kick, which will transfer to a transverse offset when the bunch returns to the cavity after recirculation. This offset leads to a change of the HOM energy. If it happens to increase the HOM energy and similar circumstance happens for the following bunches, the HOM energy will continuously grow and the beam may become unstable.

In an ERL-FEL two-purpose machine, due to the interaction of the FEL beam with the HOM field, the BBU effect becomes more complicated. If the BBU threshold current is significantly reduced due to the introduction of the FEL beam, the machine will have poor performance and become less meaningful. Thus, in this paper, taking the IHEP ERL-TF as an example, we extensively studied the BBU effect in an ERL-FEL two-purpose machine, based on previous efforts [4-5]. We developed a numerical code which considers both the ERL and FEL bunches, and calculated the BBU threshold. We found that even in the case of high FEL bunch charge, the threshold has only a small reduction (less than 5%). However, with current below the BBU threshold, we observed a fluctuation of the central orbit of the ERL-bunches in the presence of FEL beam, and theoretically clarified the cause of the orbit-fluctuation.

The paper is arranged as follows. In Sec. II, we introduce the developed BBU simulation code, and verify it with other code [7] in the case of only ERL beam; we then show the simulation results for IHEP ERL-FEL TF, including the threshold calculation and the orbit-fluctuation phenomenon. In Sec. III, we present an expanded BBU model, with which we analyze the mechanism of the orbit fluctuation induced by the FEL beam and discuss the dependence of the fluctuation amplitude to the ratio of HOM frequency versus FEL repetition rate. Conclusions are given in Sec. IV.

## 2. Simulation results

To investigate the BBU effect in an ERL-FEL two-purpose machine, we developed a new simulation code in Matlab environment based on an existing code [8]. In the simulation, both the ERL and FEL bunches are treated as macro-particles (one bunch as one macro-particle) and the wake field deflections are considered to be a point kick at the entrance of each cavity. All bunches are lined up according to their arrival time at the injection point and the coordinates of each bunch are calculated element by element. At each cavity, the transverse momentum of each bunch is updated by counting the transverse kick of the HOMs, the voltage of the HOMs is updated if the bunch has non-zero transverse offset, and the change of the bunch velocity due to acceleration or deceleration is also considered. In addition, the energy and coordinates of the bunches as well as the HOM voltage of each cavity are recorded. The obtained data are used to determine the BBU threshold at which both the HOM voltage and the transverse offset with respect to the bunch index (or the time) start increasing exponentially. To demonstrate the accuracy of the new code, for a simplest ERL model, i.e. with one cavity, one HOM, we compared the threshold current calculation results of the new code with those from theoretical analysis [6] and another code [8], as shown in Fig. 3. The agreement is very good.

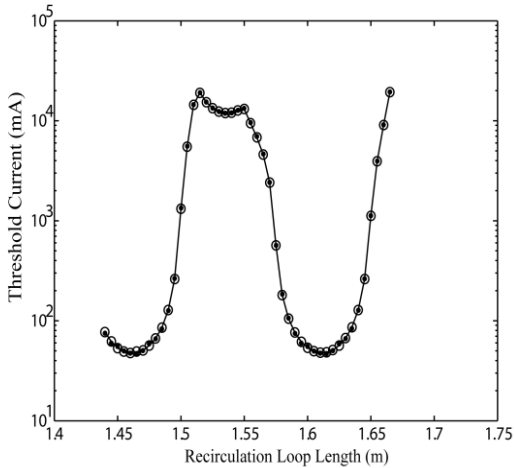

Fig. 3. Comparison of the BBU threshold calculation results of our code (dots) with those from theoretical analysis (solid line) and another code (circles) for a simplest BBU model.

Based on the verification of the code, we simulated the BBU effect in the IHEP, ERL-TF.

We first treated the case with the ERL bunches only, and then included the FEL bunches in the simulation for comparison. In the simulation we used HOMs of a TESLA 9-cell superconducting structure[9], as listed in Table 2.

Table 2. TESLA 9-cell cavity HOMs parameters

| Frequency (GHz) | Loss factor (V/pC/m$^2$) | R/Q ($\Omega$/cm$^2$) | Q ($10^4$) |
|---|---|---|---|
| 1.7949 | 21.70 | 0.77 | 1.0 |
| 1.8342 | 13.28 | 0.46 | 5.0 |
| 1.8509 | 11.26 | 0.39 | 2.5 |
| 1.8643 | 191.56 | 6.54 | 5.0 |
| 1.8731 | 255.71 | 8.69 | 7.0 |
| 1.8795 | 20.80 | 1.72 | 10 |

When only ERL bunches were considered, the threshold current of the IHEP ERL-TF was found to be about 850mA, which is much larger than the designed ERL current of 10 mA. Figure 4 shows the HOM voltage of the second cavity experienced by different ERL bunches in their first loop and the transverse position of these injected ERL bunches at the entrance of the radiator. In the case that the ERL current is sufficiently below the threshold, the voltage reaches an equilibrium value after enough time, accordingly the bunches experience the same kicks in cavities and have the same trajectories. By contrast, in the case that the current is beyond the threshold, the HOM voltages as well as the bunch coordinates grow exponentially with time and instability happens.

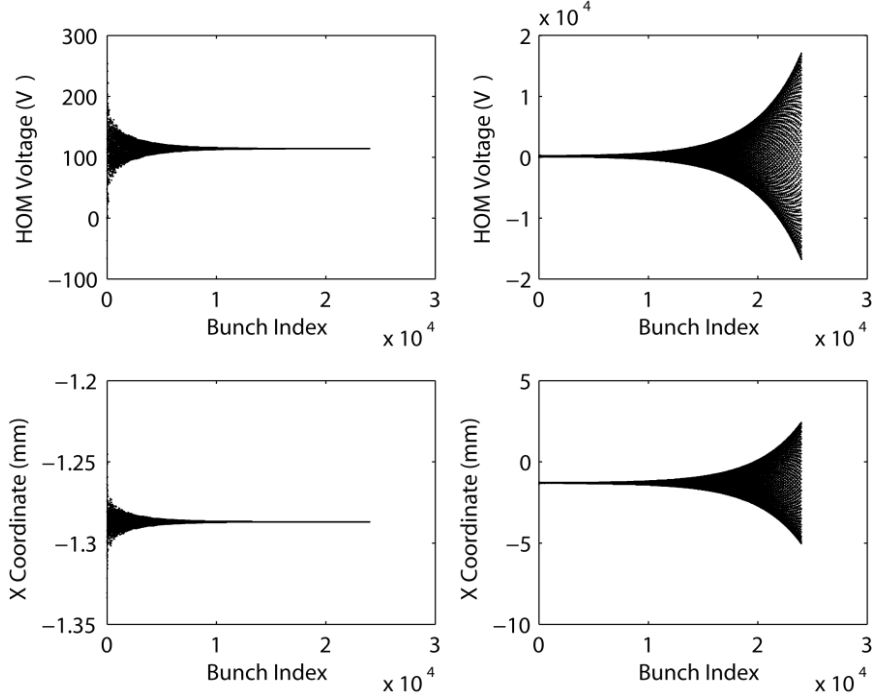

Fig. 4. the HOM voltage (above) in the second cavity experienced by different ERL bunches in their first loop and the transverse coordinates (below) of these injected ERL bunches at the end of the cavities for ERL currents 700mA, below threshold (left) and 1A, above threshold (right).

When the FEL bunches are considered, the threshold decreases gradually with increasing

FEL bunch charge, as shown in Fig.5. Fortunately, in IHEP ERL-TF this effect is not serious due to the much lower repetition rate of the FEL bunches. Even for a FEL bunch charge as large as 1 nC (corresponding to an average current of about 10mA), which is much higher than the designed FEL bunch charge of IHEP ERL-TF (100 pC), the threshold current of the facility is only reduced by about 5%.

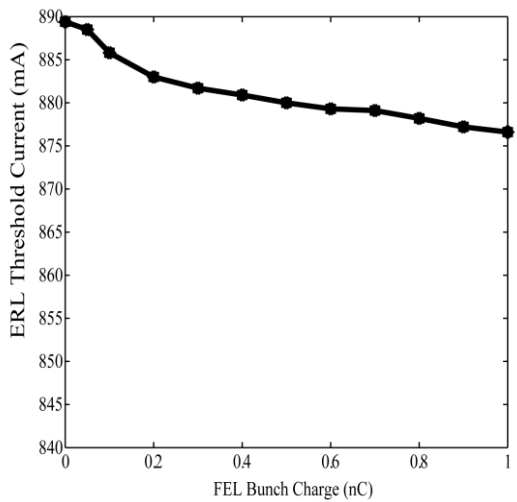

Fig. 5. The threshold current for different FEL bunch charges

Thus, in IHEP ERL-TF we don't need to worry about the threshold decrease due to the interaction of the FEL bunch. However, even for ERL beam current sufficiently below the threshold, we observed that the beam quality could be affected by the wake-field in the accelerating cavities. For illustration, the HOM voltage of the last cavity in the IHEP ERL-TF with an ERL beam of 100 mA and FEL bunch charge of 100pC is shown in Fig. 6. One can see that after thousands of ERL bunches, the HOM voltage reaches an equilibrium state. However, unlike the case with only ERL bunches (see Fig. 4), the equilibrium HOM voltage has several fixed values, instead of one. As a result, different ERL bunches will experience different kicks when they pass through the cavities. This leads to a fluctuation of the central orbits of the ERL bunches in the machine, especially in the radiator. Note that the radiation from the radiator can be optimized for only one central orbit. The orbit-fluctuation, if too large, will inevitably affect the stability of the radiation and thus the whole machine performance. We calculated the magnitude of the central orbit fluctuation, i.e. the difference between the maximum and minimum ERL bunch offset after the cavities, and used it to represent the influence of BBU effect (in presence of FEL beam) on the ERL beam quality. Fortunately, for IHEP ERL-TF with nominal beam current, the amplitude of the orbit fluctuation is on the level of μm, much smaller than the transverse size of the ERL bunch (about 1.5 mm). Therefore this effect is negligible. However, for a high energy ERL facility with beam of several GeV, the transverse size of the ERL bunch will be much smaller, and the effect of the orbit fluctuation may become important. Thus, in the next section, we formulate and clarify the mechanism of the orbit fluctuation with an expanded BBU model.

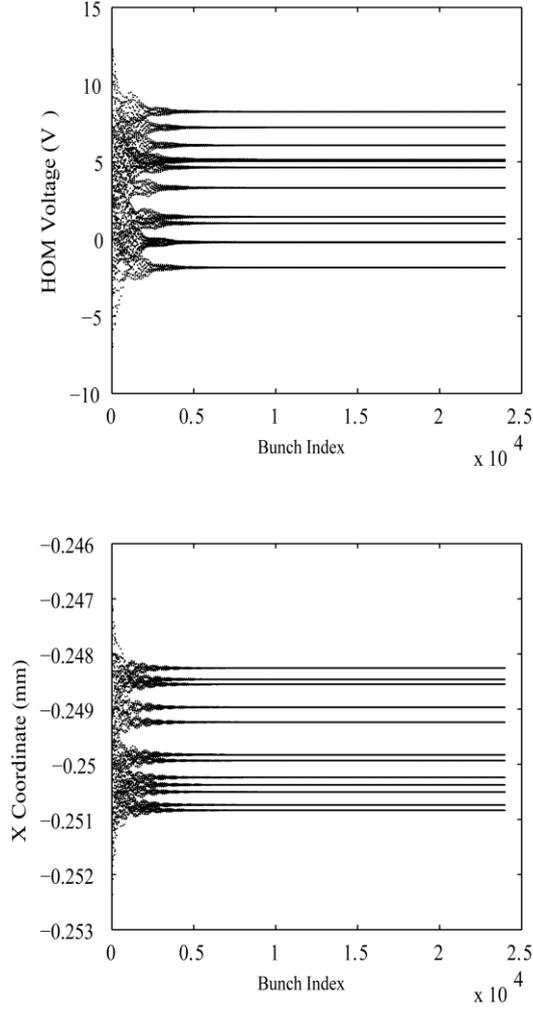

Fig. 6. HOM voltages of the last cavity (above) and the transverse position of bunches at the end of the last cavity of IHEP ERL-TF with ERL beam current of 100mA and FEL bunch charge 100 pC.

## 3. Physical model and discussions

In this section, we will analyze the interactions between HOMs and different bunches using a simplest BBU model of an ERL-FEL two-purpose machine, i.e. with one cavity and one HOM. Let us first consider that the ERL beam current is sufficiently below the BBU threshold and the FEL bunch charge is zero, after enough time the HOM voltage will reach an equilibrium value $V_0$. From the periodic condition, i.e. the HOM voltage experienced by an injected ERL bunch equals to that experienced by the injected ERL bunch after it, $V_0$ can be evaluated by:

$$V_0^C \exp[(i\omega - \frac{\omega}{2Q})T_{ERL}] + \{\exp[(i\omega - \frac{\omega}{2Q})T_{ERL}]X_{ERL} + \exp[(i\omega - \frac{\omega}{2Q})T']X_{re}\}\frac{R}{Q}\frac{\omega^2}{2c}Q_{ERL} = V_0^C,$$

$$X_{re} = M_{11}X_{ERL} + M_{12}(X'_{ERL} + \frac{eV_0}{E_{ERL}}) \quad , \tag{1}$$

where $V_0^C$ is a complex number and $V_0$ is the imaginary part of this value; $\omega$, Q and R/Q are the frequency, quality factor and shunt impedance of the HOM respectively; $T_{ERL}$ is the period of the injected ERL bunches and T' is the time between an injected ERL bunch and the recirculated ERL bunch right before it; $Q_{ERL}$ is the ERL bunch charge, $X_{ERL}$ and $X_{re}$ are the transverse position of the injected and recirculated bunch at the entrance of the cavity respectively; $M_{11}$ and $M_{12}$ are the components of the transfer matrix of recirculation loop; $X'_{ERL}$ is s derivative of X at the entrance of the cavity; c is the speed of light; $E_{ERL}$ is the energy of injected ERL beam measured in eV. In Eq. (1) the third and second terms left hand denote the HOM voltage produced by the injected bunch and the recirculated bunch right before the bunch we are interested in, and the first term is the HOM voltage produced by other bunches before it.

Considering the FEL beam whose bunch charge is not very large, the effect of FEL bunches can be considered as a perturbation on the equilibrium HOM voltage. Thus the voltage experienced by an ERL bunch can be written as a sum of $V_0$ and all voltages produced by FEL bunches before it:

$$V = V_0 + \delta V + \frac{R}{Q}\frac{\omega^2}{2c}q_{FEL}x_{FEL}\{\exp(-\frac{\omega}{2Q}t)\sin(\omega t) + \exp[-\frac{\omega}{2Q}(t+T_{FEL})]\sin[\omega(t+T_{FEL})] + \exp[-\frac{\omega}{2Q}(2t+T_{FEL})]\sin[\omega(\cdot)\cdot\cdot\cdot, \quad (2)$$

where $t$ is the time between the ERL bunch and the nearest FEL bunch before it, $q_{FEL}$, $x_{FEL}$ and $T_{FEL}$ are the bunch charge, transverse position and repetition period of the FEL bunches, respectively, $\delta V$ is a small voltage shift as a result of the interaction of FEL bunches.

In an ERL-FEL two-purpose facility, the damping time of the HOM field is much larger than the period of the FEL bunches,. Therefore, in the calculation we must consider large numbers of FEL bunches. We add up the right side of Eq. (2) as an infinite series and get Eq. (3) by taking limit

$$V = V_0 + \delta V + \frac{R}{Q}\frac{\omega^2}{2c}q_0 x_0 \operatorname{Im}\left(\frac{e^{(-\frac{\omega}{2Q}+i\omega)t}}{1-e^{(-\frac{\omega}{2Q}+i\omega)T_{FEL}}}\right). \quad (3)$$

By setting appropriate $t$ in Eq.(3), we can get all the equilibrium HOM voltages in the simplest BBU model of an ERL-FEL two-purpose facility. The calculation result was compared with the simulation in Fig. 7. Thus the mechanism of the equilibrium voltage spread can be understood as a result of a perturbation introduced by the FEL bunches. What's more, in Eq. (3) the magnitude of HOM voltage is a function of the frequency of HOM field, and we can make an assumption of the voltage split as:

$$\Delta V = \frac{R}{Q}\frac{\omega^2}{2c}q_0 x_0 \left\{\max\left[\operatorname{Im}\left(\frac{e^{(-\frac{\omega}{2Q}+i\omega)t}}{1-e^{(-\frac{\omega}{2Q}+i\omega)T_{FEL}}}\right)\right] - \min\left[\operatorname{Im}\left(\frac{e^{(-\frac{\omega}{2Q}+i\omega)t}}{1-e^{(-\frac{\omega}{2Q}+i\omega)T_{FEL}}}\right)\right]\right\}. \quad (4)$$

When $\omega T_{FEL} = 2n\pi$, a resonance occurs in Eq. (4) and the voltage split get the maximum value. For the simplest BBU model (one cavity, one HOM), we compared the simulation results and those from Eq. (4) for different FEL repetition rate in Fig. 8. We can see that when the HOM frequency is an integer multiple of the FEL repetition rate, the magnitude of the voltage split can be much larger than that in other cases. From a simulation result of the HOM voltage spread and orbit fluctuation for different FEL frequencies in the IHEP ERL-TF shown in Fig.9, this resonance relationship, which is revealed by Eq. (3), also exists in more complicated machines. Thus, in the design of a two-purpose machine, the central orbit fluctuation can be minimized by an appropriate choice of the FEL repetition rate.

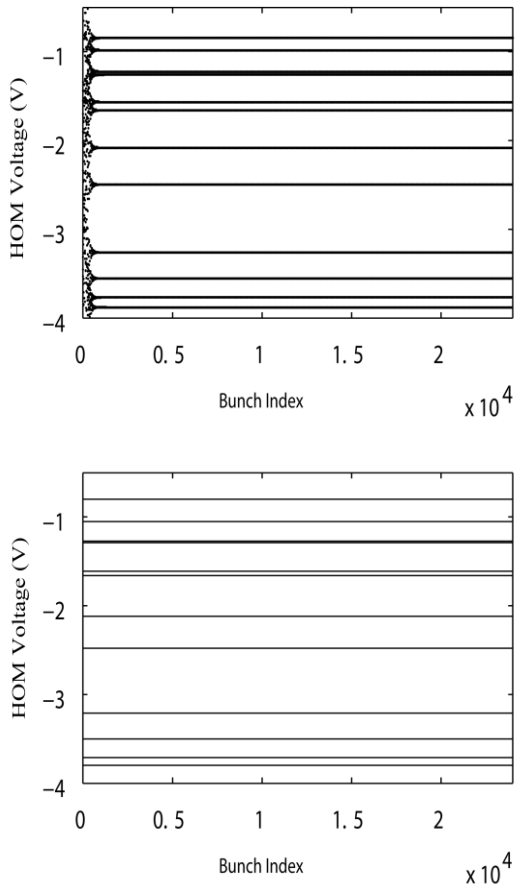

Fig. 7. Comparison of the equilibrium HOM voltage in the cavity from simulation (above) with that calculated by Eq. (3), while neglecting δV (below).

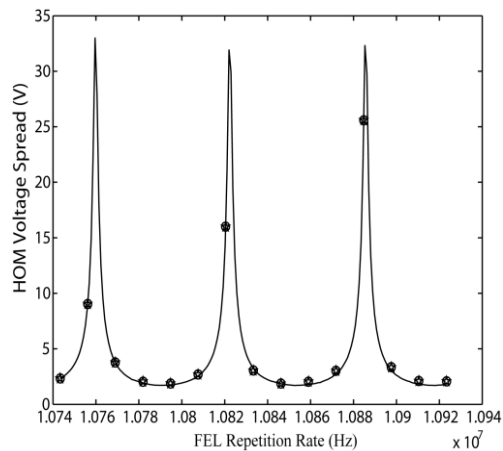

Fig. 8. Comparison of the magnitude of the voltage split for different FEL repetition rate from simulation (dots) and that calculated from Eq. (4) (line).

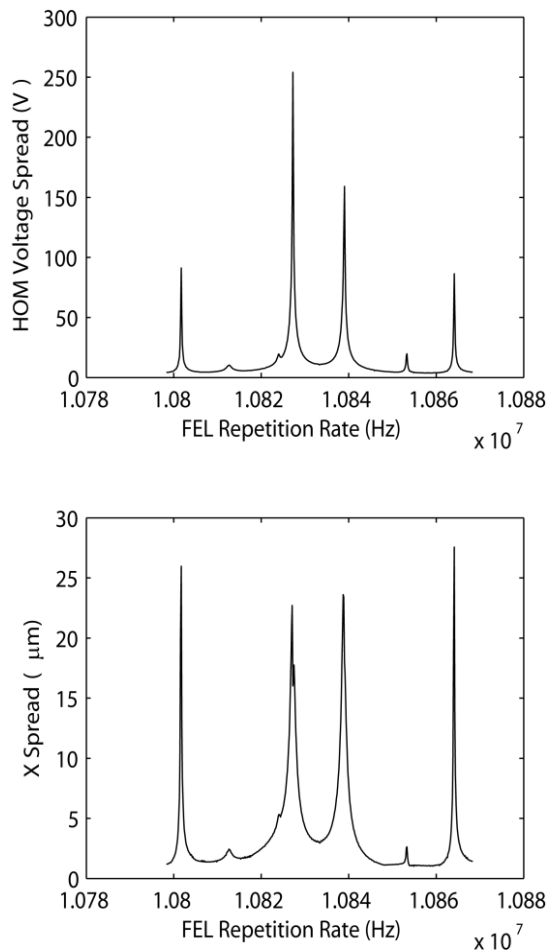

Fig. 9. magnitude of the HOM voltage spread (above) and the ERL bunch orbit fluctuation (below) for different FEL frequencies.

## 4. Conclusions

In this paper, taking IHEP ERL-TF as an example, we studied the BBU effect in an ERL-FEL two-purpose machine. We found that two effects emerge as a result of the introduction of FEL beams: a reduction in the threshold current and a central orbit fluctuation for ERL current under threshold. Due to the fact that the repetition rate of FEL bunches is much smaller than that of ERL, the introduction of FEL beam should not have a fatal effect on the threshold current. As for the orbit fluctuation, we gave a simple model and found a resonance relation between the voltage spread and the ratio of HOM frequency to the FEL repetition rate. By choosing an appropriate FEL frequency, the amplitude of the orbit fluctuation can be kept small.